\documentclass[%
 aip,
rsi,%
 amsmath,amssymb,
 reprint,%
longbibliography,
]{revtex4-1}

\usepackage{xcolor, soul}
\sethlcolor{yellow}
\usepackage{graphicx}% Include figure files
\usepackage{dcolumn}% Align table columns on decimal point
\usepackage{bm}% bold math
\usepackage{euscript}
\usepackage{enumitem}
\usepackage[colorlinks=true,bookmarks=false,citecolor=blue,urlcolor=blue]{hyperref}     
\usepackage{booktabs,siunitx}
\usepackage{multirow}

\begin{document}

\preprint{AIP/123-QED}

\title[Hybrid Plasmon--Phonon Eigenmodes and Vortex States]{Hybrid Plasmon–Phonon Eigenmodes and Vortex States in a Graphene Ring on a Silica Substrate}
% Force line breaks with \\
%\thanks{Footnote to title of article.}

\author{Yanxu~Guo}
\affiliation{State Key Laboratory on Integrated Optoelectronics, College of Electronic Science and Engineering, International Center of Future Science, Jilin University, 2699 Qianjin Street, Changchun 130012, China} 
\author{Oleh~Demianyk}
\affiliation{School of Radiophysics, Biomedical Electronics and Computer Systems, V.~N.~Karazin Kharkiv National University, 4, Svobody Square, Kharkiv 61022, Ukraine} 
\author{Vladimir~R.~Tuz}
\affiliation{State Key Laboratory on Integrated Optoelectronics, College of Electronic Science and Engineering, International Center of Future Science, Jilin University, 2699 Qianjin Street, Changchun 130012, China} 
\affiliation{School of Radiophysics, Biomedical Electronics and Computer Systems, V.~N.~Karazin Kharkiv National University, 4, Svobody Square, Kharkiv 61022, Ukraine}

\date{\today}% It is always \today, today,
             % but any date may be explicitly specified

\begin{abstract}
We investigate hybrid plasmon--phonon eigenmodes and localized electromagnetic vortex states in a graphene ring supported by a dispersive silica substrate. The interaction between graphene plasmons and infrared-active optical phonons of the substrate gives rise to distinct spectral branches of surface polaritons, whose dispersion characteristics are established analytically and validated by full-wave numerical simulations. Geometric confinement of these hybrid surface excitations produces spectrally separated families of localized multipolar eigenmodes with corresponding azimuthal symmetries. Particular attention is devoted to quadrupolar eigenstates, which exhibit nearly identical normalized spatial field distributions despite their substantial spectral separation and different attenuation characteristics. We demonstrate that coherent superposition of orthogonal quadrupolar field patterns gives rise to localized vortex states with opposite azimuthal phase winding. The preservation of the spatial eigenmode structure across distinct spectral branches is consistent with a quasistatic description in which the resonator geometry governs the modal symmetry, whereas the dispersive material response determines the spectral properties. These findings reveal a route toward controlling structured electromagnetic near fields through plasmon--phonon hybridization and establish a physical basis for multiband vortex states in deeply subwavelength graphene resonators.
\end{abstract}

%Valid PACS numbers may be entered using the \verb+\pacs{#1}+ command.

%\pacs{Valid PACS appear here}% PACS, the Physics and Astronomy Classification Scheme.

%\keywords{Metasurface, electromagnetic induced transparency, patch resonator, slow light effects, group delay}
%Use showkeys class option if keyword display desired

\maketitle

\section{Introduction}
\label{sec:intro}

Structured electromagnetic fields carrying orbital angular momentum (OAM) have attracted considerable attention owing to their distinctive spatial phase distributions and their ability to introduce additional degrees of freedom for controlling light--matter interactions.\cite{yao_2011_AOP, bliokh_2015_PR, bliokh_2015_NP, Roadmap_2017} Optical vortex states are commonly characterized by an azimuthal phase dependence $\exp(il\varphi)$, where $l$ is an integer topological charge describing the phase winding around the vortex axis.\cite{Barnett_rsta_2017} Such states provide opportunities for encoding information in spatial modes, manipulating microscopic objects, and engineering the transfer of angular momentum between electromagnetic fields and matter. \cite{Gibson_OE_2004, yang_2021_AP, Willner_APL_2021, willner_2023_JLT} Beyond freely propagating vortex beams, resonant photonic systems can also support structured electromagnetic fields, including optical nanoantennas, plasmonic nanostructures, and metasurfaces. At subwavelength scales, the confinement of electromagnetic energy and the formation of spatially varying phase distributions offer additional possibilities for controlling structured optical near fields.

Optical nanoantennas provide a versatile platform for manipulating electromagnetic fields at spatial scales considerably smaller than the free-space wavelength.\cite{Novotny_2011, Giannini_2011, Biagioni_2012} Their resonant response enables strong field localization and enhanced light--matter interactions, making them attractive building blocks for nanophotonic devices, nonlinear optical systems, and structured-light generation.\cite{Fischer_2008, Bakker_2008, Esteban_2010, Chen_2012, Hoang_2015, Fitzgerald_2017} Of particular interest are rotationally symmetric resonators, whose eigenmodes can be classified according to their azimuthal order. In circular and ring-shaped structures, pairs of orthogonal standing-wave modes can be combined with an appropriate relative phase to form azimuthally propagating waves with a well-defined phase winding.\cite{Muelas-Hurtado_APL_2018, Mao_2018} This principle has been employed in ring-shaped resonator antennas to generate electromagnetic fields carrying different OAM states and to implement multiband mode-division multiplexing.\cite{He_OptExpress_2022, He_APL_2023} Extending this approach to deeply subwavelength plasmonic resonators offers a route toward structured electromagnetic fields confined to nanoscale dimensions.\cite{Sahoo_PRA_2026}

Graphene is particularly attractive for such applications because it supports strongly confined surface plasmon polaritons (SPPs) in the terahertz and mid-infrared spectral regions.\cite{Grigorenko_2012,Jablan_2013,Luo_2013,Politano_2014,Abajo_2014} Unlike conventional plasmonic metals, graphene exhibits a frequency-dependent surface conductivity that can be modified by varying the carrier concentration, temperature, or electrostatic bias. This property enables control over plasmonic dispersion, field confinement, and resonant frequencies, providing a basis for actively tunable nanophotonic structures.\cite{Koppens_2011,Filter_2013} Graphene-based resonators, including ribbons, disks, and ring-shaped geometries, have been investigated as compact building blocks for terahertz and infrared nanoantennas and metasurfaces.\cite{Liu_2012,Llatser_2012,Tamagnone_2012,Liu_2014,Nissiyah_2019} In particular, the rotational symmetry of a graphene ring permits the formation of dipolar, quadrupolar, and higher-order plasmonic eigenmodes, whose azimuthal field distributions provide a natural basis for constructing localized vortex states.

An additional degree of control over graphene plasmons can be introduced through their interaction with the dielectric environment. When graphene is supported by a polar dielectric substrate, collective charge-carrier oscillations can interact with infrared-active optical phonons, giving rise to hybrid plasmon--phonon polaritons.\cite{Brar_nl_2014, Jia_acsphotonics_2015} The resulting electromagnetic response can differ substantially from that of graphene on a nondispersive dielectric, particularly near the phonon resonances where the substrate permittivity exhibits rapid frequency variations and may become negative. Such hybridization modifies the surface-wave dispersion, confinement, and attenuation, potentially producing distinct spectral branches with different propagation characteristics. For a finite graphene resonator, this dispersive response provides an additional mechanism for controlling the frequencies and damping of localized multipolar eigenmodes. The possibility of combining substrate-induced spectral hybridization with the azimuthal mode structure of a graphene ring is especially relevant to the realization of structured plasmonic near fields in multiple infrared frequency bands.

Although the plasmonic properties of graphene nanostructures and the generation of OAM states by circular resonators have been investigated extensively, the relationship between the hybrid plasmon--phonon dispersion of a graphene--polar-dielectric interface and the formation of localized vortex-supporting eigenmodes in a finite graphene ring remains an important problem. In particular, it is of interest to determine how the frequency-dependent substrate response influences the ordering, spatial symmetry, and damping of the ring eigenmodes, and whether modes of the same azimuthal order can occur in separated spectral regions. Quadrupolar resonances are especially attractive in this context because their orthogonal standing-wave components can be combined into counter-rotating azimuthal states characterized by phase factors $\exp(\pm 2i\varphi)$. Establishing such a connection between surface-wave dispersion, localized eigenmodes, and azimuthal phase winding would provide a physical basis for multiband structured-field excitation in compact graphene nanostructures.

In this work, we investigate the dispersion and localized electromagnetic eigenmodes of a graphene ring deposited on a dispersive silica (SiO$_2$) substrate, with particular emphasis on their suitability for generating quadrupolar vortex states. We first derive the dispersion relation for TM-polarized SPPs at the planar graphene--SiO$_2$ interface and examine how the resonant phonon response of the substrate modifies their propagation characteristics. The analytical results are compared with full-wave numerical simulations, revealing two separated spectral branches associated with hybrid plasmon--phonon polaritons. We formulate an effective surface-wave model and a quasistatic integral eigenvalue problem to establish the theoretical basis for classifying the localized resonances of the graphene ring. The complex eigenfrequencies and corresponding electromagnetic field distributions are subsequently determined using full-wave finite-element simulations. Numerical eigenmode calculations identify dipolar, quadrupolar, and hexapolar field distributions, occurring in both spectral regions. Particular attention is paid to the differences in modal damping and to the possibility of constructing vortex states with opposite azimuthal phase winding through the coherent superposition of orthogonal quadrupolar modes. The proposed framework connects hybrid plasmon--phonon eigenmode formation with structured electromagnetic near fields and provides a basis for investigating multiband vortex excitation in deeply subwavelength graphene resonators.

\section{Materials Description}
\label{sec:materials}

The electromagnetic properties of the graphene--SiO$_2$ structure are determined by the frequency-dependent surface conductivity of graphene and the dispersive dielectric response of the SiO$_2$ substrate. Both materials exhibit distinct physical mechanisms that influence the propagation and confinement of surface electromagnetic waves. In particular, the interaction between collective charge-carrier oscillations in graphene and infrared-active optical phonons in SiO$_2$ provides the basis for the formation of hybrid plasmon--phonon polaritons.

\subsection{Graphene conductivity}

Neglecting nonlocal effects and quantum finite-size corrections, graphene is described by a complex surface conductivity $\sigma(\omega)$, which depends on the angular frequency $\omega=2\pi f$, Fermi energy $E_F$, temperature $T$, and carrier relaxation time $\tau$.

Within the local Kubo formalism, \cite{Falkovsky_PhysRev_2007} the graphene surface conductivity is expressed as the sum of intraband and interband contributions,
\begin{equation}
\sigma(\omega)=
\sigma_{\mathrm{intra}}(\omega)+
\sigma_{\mathrm{inter}}(\omega).
\label{eq:graphene_sigma}
\end{equation}
The intraband contribution describes the response of free charge carriers, whereas the interband contribution accounts for electronic transitions between the valence and conduction bands.

For the harmonic time dependence $\exp(+i\omega t)$ adopted throughout this work, the intraband and interband contributions are given by
\begin{equation}
\begin{split}
\sigma_{\mathrm{intra}} &= -\frac{ie^2 k_B T}{\pi\hbar^2(\omega-i\Gamma)}\left[\frac{E_F}{k_B T} + 2\ln\left(1+e^{-\frac{E_F}{k_B T}}\right)\right],\\
\sigma_{\mathrm{inter}}
&=
-\frac{ie^2}{4\pi\hbar}\ln\left[\frac{2|E_F| - \hbar(\omega-i\Gamma)}{2|E_F| + \hbar(\omega-i\Gamma)}\right].
\end{split}
\label{eq:Kubo}
\end{equation}
Here, $e$ is the elementary charge, $k_B$ is the Boltzmann constant, and $\hbar$ is the reduced Planck constant. The Fermi energy $E_F$ is expressed in joules, and the phenomenological broadening parameter is defined as $\Gamma=2/\tau$, where $\tau$ is the carrier relaxation time, consistently with the conductivity model implemented in COMSOL. The interband contribution is described using an analytical approximation that incorporates phenomenological collision broadening.

The calculations are performed for $E_F=0.3$~eV, $T=300$~K, and $\tau=6.5\times10^{-14}$~s. In the finite-element model, graphene is represented by an effective conducting layer with thickness $d_{\mathrm{eff}}=0.345$~nm. Its equivalent bulk electrical conductivity is defined as
\begin{equation}
\sigma_{\mathrm{3D}}(\omega)
=
\frac{\sigma(\omega)}{d_{\mathrm{eff}}}.
\label{eq:graphene_bulk_sigma}
\end{equation}
For the adopted time convention, the corresponding effective relative permittivity can be written as
\begin{equation}
\varepsilon_g(\omega)
=
1-\frac{i\sigma(\omega)}
{\varepsilon_0\omega d_{\mathrm{eff}}},
\label{eq:graphene_eps}
\end{equation}
where $\varepsilon_0$ is the vacuum permittivity.

In the analytical treatment, graphene is represented by an infinitesimally thin conducting sheet with surface conductivity $\sigma(\omega)$, whereas the finite-element simulations employ an effective conducting layer of thickness $d_{\mathrm{eff}}$. These descriptions become equivalent in the thin-layer limit, provided that the electromagnetic field varies negligibly across the effective graphene thickness. This condition is satisfied when $|\kappa_g|d_{\mathrm{eff}}\ll 1$.

Figure~\ref{fig:fig1}(a) shows the real and imaginary parts of the graphene surface conductivity in the considered frequency range. The real part, which characterizes dissipative losses associated with charge-carrier motion, decreases monotonically with increasing frequency. The imaginary part remains negative throughout the displayed interval and gradually approaches zero in magnitude. This behavior is characteristic of the predominantly intraband response of electron-doped graphene. The negative imaginary conductivity provides the inductive surface response required for the conventional TM-polarized graphene plasmon.

\subsection{Dielectric response of SiO$_2$}

The SiO$_2$ substrate is described by a complex, frequency-dependent relative permittivity $\varepsilon_{\mathrm{SiO_2}}(\omega)$. Unlike a nondispersive dielectric, SiO$_2$ exhibits pronounced resonant features in the infrared spectral range associated with the excitation of polar optical phonons.

The optical properties of the substrate are taken from the experimental data,\cite{Popova_1972} available in the COMSOL material database as ``SiO$_2$ (Silicon dioxide, Silica, Quartz): Popova et al. (1972), fused silica, $n,k$ at 7--50~$\mu$m''.

The tabulated real refractive index $n_r(\lambda)$ and extinction coefficient $n_i(\lambda)$ are converted into the complex relative permittivity according to
\begin{equation}
\varepsilon_{\mathrm{SiO_2}}(\omega)
=
\left[n_r(\omega)-in_i(\omega)\right]^2.
\label{eq:SiO2_eps}
\end{equation}
The wavelength and frequency are related by $\lambda=2\pi c/\omega=c/f$, where $c$ is the speed of light in vacuum.

The corresponding real and imaginary parts of the permittivity are
\begin{equation}
\begin{split}
\operatorname{Re}
\left(\varepsilon_{\mathrm{SiO_2}}\right)
&=n_r^2-n_i^2,\\
\operatorname{Im}
\left(\varepsilon_{\mathrm{SiO_2}}\right)
&=-2n_r n_i.
\end{split}
\label{eq:SiO2_eps_components}
\end{equation}

\begin{figure}[!t]
\centering
\includegraphics[width=\linewidth]{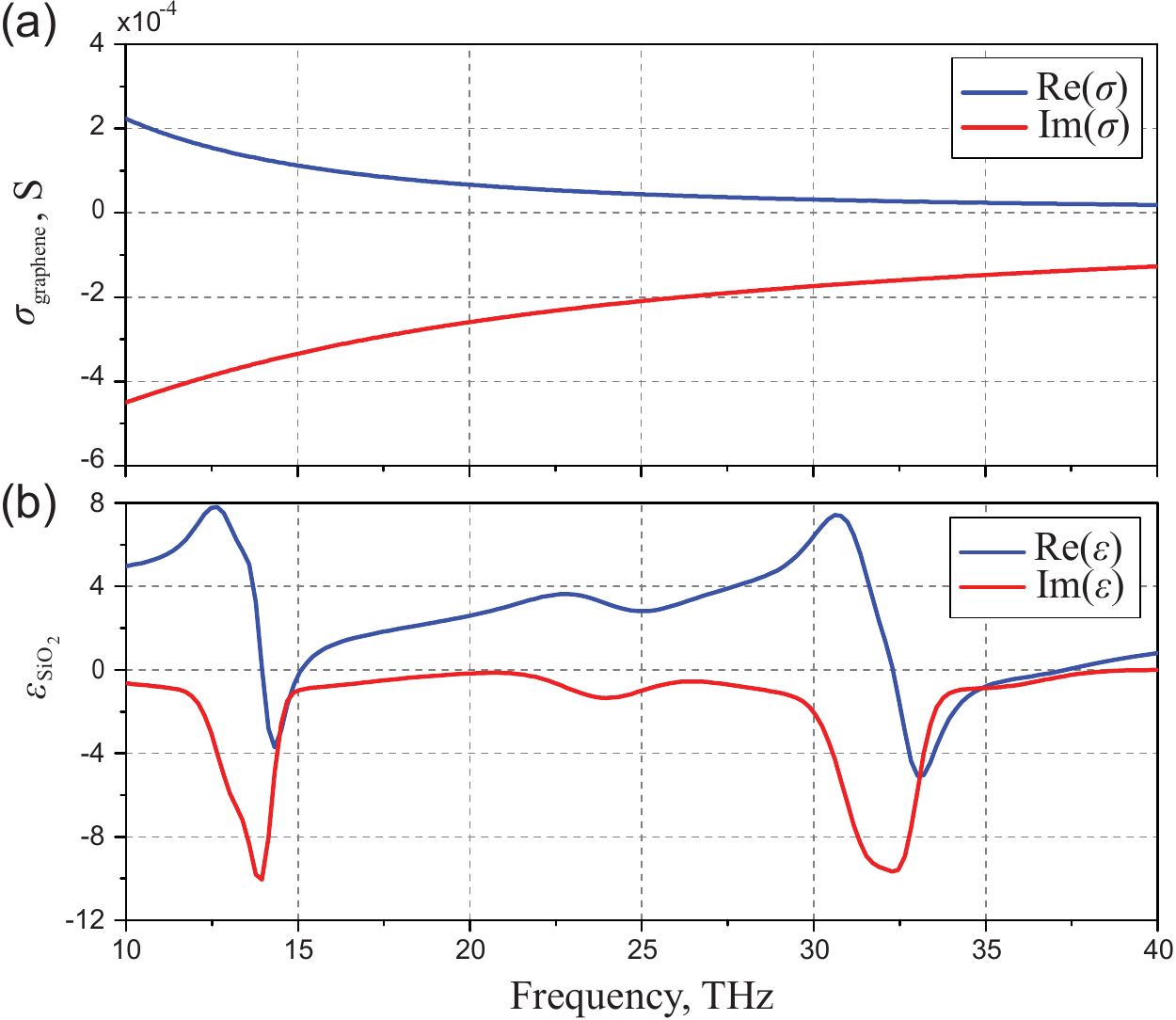}
\caption{Frequency dependence of the complex electromagnetic properties of graphene and SiO$_2$. (a) Real (blue) and imaginary (red) parts of the graphene surface conductivity $\sigma(\omega)$, calculated using the Kubo model for $E_F=0.3$~eV, $T=300$~K, and $\tau=65$~fs. (b) Real (blue) and imaginary (red) parts of the relative permittivity $\varepsilon_{\mathrm{SiO_2}}(\omega)$, calculated from the experimental refractive-index and extinction-coefficient data\cite{Popova_1972} for fused silica. The pronounced resonances originate from infrared-active optical phonons and produce frequency intervals with negative real permittivity.}
\label{fig:fig1}
\end{figure}

Figure~\ref{fig:fig1}(b) presents the frequency dependence of the complex relative permittivity of SiO$_2$. Two prominent resonant regions are observed, approximately at 13--15~THz and 30--33~THz. In the vicinity of these resonances, the real part of the permittivity undergoes rapid variations and becomes negative over certain frequency intervals, while the imaginary part exhibits pronounced negative peaks corresponding to strong optical absorption.

These features originate from infrared-active optical phonons of the SiO$_2$ substrate. The regions of negative real permittivity are particularly important because they can support surface phonon polaritons at an interface with a dielectric of positive permittivity.

In the graphene--SiO$_2$ structure, the electromagnetic fields of graphene plasmons interact with the resonant polarization response of the substrate. This interaction modifies the dispersion, spatial confinement, and attenuation of the surface modes. Consequently, the hybrid plasmon--phonon response can exhibit pronounced dispersion bending and separated frequency branches in the vicinity of the phonon resonances.

To distinguish these substrate-induced features from the intrinsic dispersion of graphene plasmons, a reference structure with a nondispersive dielectric substrate of constant relative permittivity $\varepsilon_d=2.1$ is also considered. This reference model retains the same graphene conductivity but excludes the resonant phonon response of SiO$_2$, providing a baseline for identifying the effects of plasmon--phonon coupling.

\section{SPP Dispersion Equation}
\label{sec:spp_dispersion}

\subsection{Analytical Formulation of TM Surface Modes}
\label{subsec:spp_analytical}

We derive the dispersion relation for electromagnetic surface modes supported by a graphene sheet located at the planar interface between two semi-infinite dielectric media. The analytical formulation treats graphene as an infinitesimally thin conducting sheet characterized by the frequency-dependent surface conductivity $\sigma(\omega)$ introduced in Sec.~\ref{sec:materials}.

We consider two dielectric half-spaces separated by the plane $z=0$. The upper medium ($z>0$) is characterized by a relative permittivity $\varepsilon_1$, whereas the lower medium ($z<0$) has a relative permittivity $\varepsilon_2$. Both media are assumed to be homogeneous, isotropic, and nonmagnetic.

We seek surface electromagnetic waves propagating along the $x$-direction with an in-plane propagation constant $q$ and fields that decay exponentially away from the interface. In the following, we focus on transverse-magnetic (TM) surface modes, which constitute the conventional graphene surface plasmon polaritons (SPPs). For this polarization, the nonzero electromagnetic field components are $E_x$, $E_z$, and $H_y$. The spatial dependence along the interface is taken as $\exp(iqx)$, while the dependence along the $z$-direction is evanescent.

For the adopted harmonic time dependence $\exp(+i\omega t)$, Maxwell's equations in each dielectric region read
\begin{align}
\nabla\times\mathbf{E}
&=-i\omega\mu_0\mathbf{H},\\
\nabla\times\mathbf{H}
&=i\omega\varepsilon_0\varepsilon_j\mathbf{E},
\qquad j=1,2.
\end{align}

In the upper region $z>0$ (medium 1), the electromagnetic fields are expressed as
\begin{align}
E_x^{(1)}
&=A_1 e^{iqx-\kappa_1 z},\\
E_z^{(1)}
&=\frac{iq}{\kappa_1}
A_1 e^{iqx-\kappa_1 z},\\
H_y^{(1)}
&=\frac{i\varepsilon_0\varepsilon_1\omega}
{\kappa_1}
A_1 e^{iqx-\kappa_1 z}.
\end{align}

In the lower region $z<0$ (medium 2), the fields take the form
\begin{align}
E_x^{(2)}
&=A_2 e^{iqx+\kappa_2 z},\\
E_z^{(2)}
&=-\frac{iq}{\kappa_2}
A_2 e^{iqx+\kappa_2 z},\\
H_y^{(2)}
&=-\frac{i\varepsilon_0\varepsilon_2\omega}
{\kappa_2}
A_2 e^{iqx+\kappa_2 z}.
\end{align}

Here, $A_1$ and $A_2$ are complex field amplitudes, and the transverse decay constants are defined as
\begin{equation}
\kappa_j=
\sqrt{q^2-\varepsilon_j k_0^2},
\qquad j=1,2,
\end{equation}
where $k_0=\omega/c$ is the free-space wavenumber
and $c=1/\sqrt{\mu_0\varepsilon_0}$ is the speed
of light in vacuum. The branches of the square
roots are chosen such that
\begin{equation}
\operatorname{Re}(\kappa_j)>0,
\qquad j=1,2,
\end{equation}
ensuring exponential localization of the fields near the graphene interface. In general, the propagation constant $q$ is complex, accounting for attenuation along the propagation direction.

The continuity of the tangential electric field at $z=0$ requires
\begin{equation}
E_x^{(1)}(0)=E_x^{(2)}(0),
\end{equation}
which yields
\begin{equation}
A_1=A_2\equiv A.
\end{equation}

The graphene sheet introduces a discontinuity in the tangential magnetic field due to the induced surface current density. With the unit normal $\hat{\mathbf z}$ directed from medium 2 to medium 1, the corresponding boundary condition is
\begin{equation}
\hat{\mathbf z}\times
\left(\mathbf H_1-\mathbf H_2\right)
=\mathbf J_s,
\end{equation}
where
\begin{equation}
\mathbf J_s=\sigma(\omega)\mathbf E_\parallel
\end{equation}
is the surface current density and $\mathbf E_\parallel$ denotes the tangential electric field at the interface.

For TM-polarized modes, the magnetic-field boundary condition reduces to
\begin{equation}
H_y^{(2)}(0)-H_y^{(1)}(0)
=\sigma(\omega)E_x(0).
\end{equation}

Substituting the field expressions and using the continuity of $E_x$, we obtain
\begin{equation}
-i\varepsilon_0\omega\left(\frac{\varepsilon_1}{\kappa_1} + \frac{\varepsilon_2}{\kappa_2}\right)A =\sigma(\omega)A.
\end{equation}

For a nontrivial surface mode ($A\neq 0$), this condition leads to the TM-mode dispersion equation
\begin{equation}
\frac{\varepsilon_1}{\kappa_1} + \frac{\varepsilon_2}{\kappa_2} - \frac{i\sigma(\omega)}{\varepsilon_0\omega} = 0.
\label{eq:graphene_dispersion_SI}
\end{equation}

Equation~\eqref{eq:graphene_dispersion_SI} determines the complex propagation constant $q(\omega)$ of TM-polarized surface electromagnetic modes supported by the graphene interface. For the graphene--SiO$_2$ structure, we set $\varepsilon_1=1$ and $\varepsilon_2=\varepsilon_{\mathrm{SiO_2}}(\omega)$, where the frequency-dependent permittivity of the substrate is defined by Eq.~\eqref{eq:SiO2_eps}.

In the nonretarded regime, $|q|\gg k_0\sqrt{|\varepsilon_j|}$, the transverse decay constants satisfy $\kappa_j\approx q$ for the strongly confined branch with $\operatorname{Re}(q)>0$. The dispersion equation then reduces to
\begin{equation}
q\approx
-\frac{i\varepsilon_0
(\varepsilon_1+\varepsilon_2)\omega}
{\sigma(\omega)}.
\label{eq:graphene_dispersion_nonretarded}
\end{equation}

For dielectric surroundings with positive real permittivities and sufficiently weak losses, the conventional TM-plasmon branch requires $\operatorname{Im}[\sigma(\omega)]<0$ under the adopted $\exp(+i\omega t)$ convention. With the spatial dependence $\exp(iqx)$, attenuation along the positive $x$-direction corresponds to $\operatorname{Im}(q)>0$.

\subsection{Dispersion Characteristics and Numerical Validation}
\label{subsec:spp_numerical}

\begin{figure}[!t]
\centering
\includegraphics[width=\linewidth]{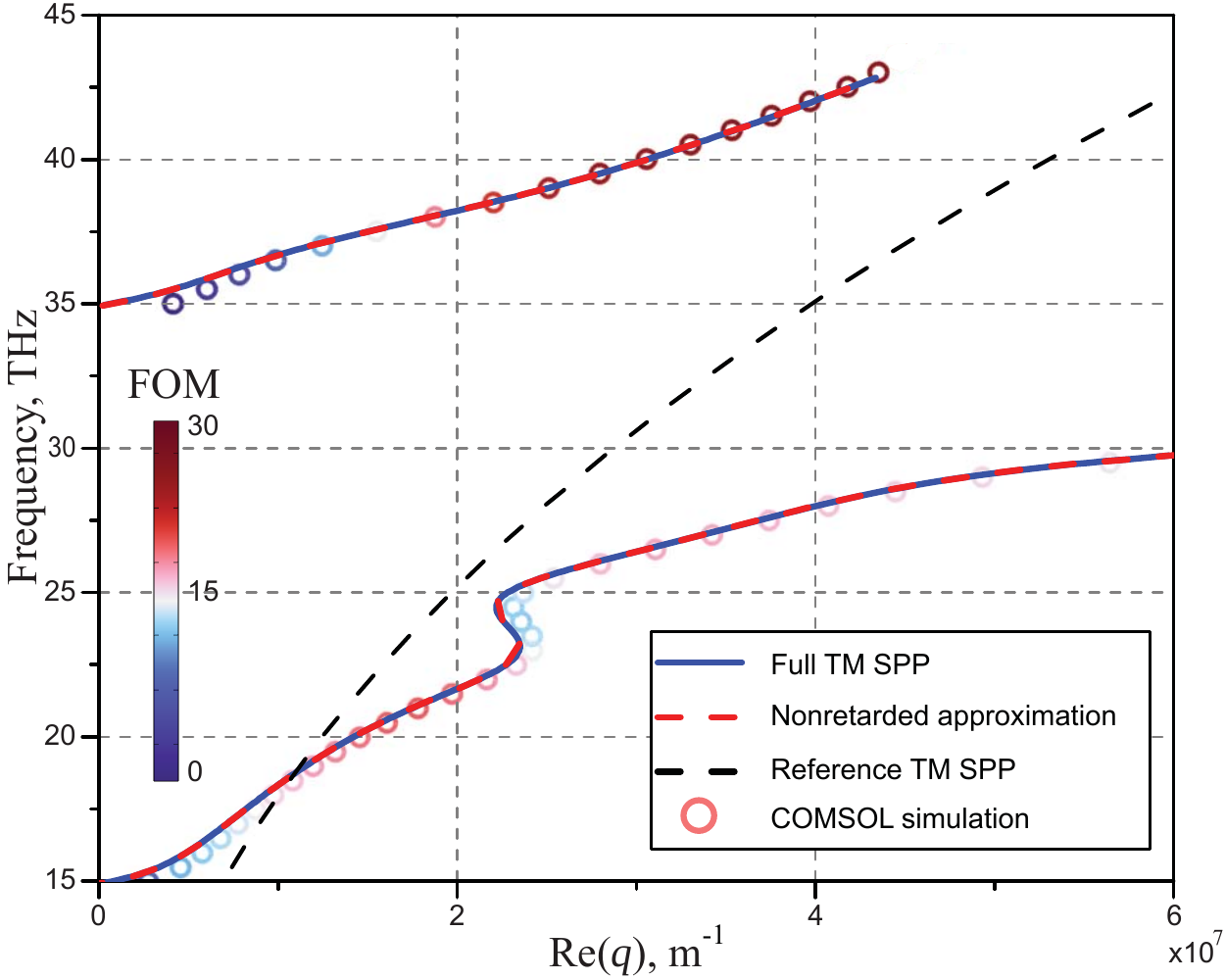}
\caption{Dispersion characteristics of TM-polarized surface plasmon polaritons in the graphene--SiO$_2$ structure. Blue solid lines represent solutions of the full retarded TM dispersion equation, while red dashed lines correspond to the nonretarded approximation. The black dashed line shows the reference TM-plasmon dispersion for a nondispersive dielectric substrate with $\varepsilon_d=2.1$. Open circles represent full-wave numerical results obtained using COMSOL Multiphysics, with their colors indicating the propagation figure of merit, $\mathrm{FOM}=\operatorname{Re}(q)/|\operatorname{Im}(q)|$. The separated spectral branches reflect the influence of the resonant dielectric response of SiO$_2$ associated with its infrared-active optical phonons. The dielectric dispersion of SiO$_2$ is based on the experimental optical constants reported in Ref.~\onlinecite{Popova_1972}, covering wavelengths of $7$--$50$~$\mu$m (approximately $6.0$--$42.8$~THz).}
\label{fig:fig2}
\end{figure}

Figure~\ref{fig:fig2} presents the dispersion characteristics of TM-polarized surface modes supported by the graphene--SiO$_2$ interface. The solutions of the full retarded dispersion equation~\eqref{eq:graphene_dispersion_SI} are compared with the nonretarded approximation given by Eq.~\eqref{eq:graphene_dispersion_nonretarded} and with full-wave numerical results obtained using COMSOL Multiphysics. The numerical model was implemented following the SPP simulation approach described in Ref.~\onlinecite{Chen_COMSOL_2022}, employing the Electromagnetic Waves, Frequency Domain interface and Boundary Mode Analysis to determine the complex propagation constants of the surface modes. The calculations use the graphene conductivity parameters and frequency-dependent dielectric response of SiO$_2$ introduced in Sec.~\ref{sec:materials}.

The analytical and numerical dispersion characteristics exhibit good agreement over the investigated frequency range. The nonretarded approximation closely reproduces the full TM dispersion over the strongly confined portions of both spectral branches. Deviations may become more pronounced where the in-plane propagation constant approaches the characteristic wavenumbers of the surrounding media, indicating the increasing importance of retardation effects.

The dispersion characteristics exhibit two separated spectral branches, approximately within the frequency intervals 15--30~THz and 35--43~THz, with the upper-frequency range limited by the available experimental dielectric data. The lower branch displays a pronounced change in curvature near 23--25 THz, followed by a substantial increase in the propagation constant as the frequency approaches 30 THz. The upper branch exhibits a comparatively smooth, monotonic increase in the propagation constant with frequency.

These features arise from the strongly dispersive dielectric response of SiO$_2$, particularly in the vicinity of its infrared-active optical phonon resonances [Fig.~\ref{fig:fig1}(b)]. As also discussed in Ref.~\onlinecite{Chen_COMSOL_2022}, the phonon-induced negative real permittivity of SiO$_2$ can suppress conventional graphene SPP propagation over a certain frequency interval, leading to separated spectral branches.

The frequency interval separating the two branches corresponds to a region where the resonant substrate response strongly modifies the conditions for surface-wave propagation. In this region, the solutions of the complex dispersion equation may exhibit enhanced attenuation or reduced field localization, limiting the existence of well-defined propagating SPP modes.

To distinguish the influence of substrate phonons from the intrinsic plasmonic response of graphene, Fig.~\ref{fig:fig2} also includes a reference dispersion curve calculated for a nondispersive dielectric substrate with a constant relative permittivity $\varepsilon_d=2.1$. This reference structure retains the same graphene conductivity and interface geometry but excludes the resonant phonon response of SiO$_2$.

Unlike the dispersive graphene--SiO$_2$ structure, the reference system exhibits a single smooth, monotonically increasing TM-plasmon branch without pronounced resonance-induced bending or spectral separation. This comparison demonstrates that the distinctive features of the graphene--SiO$_2$ dispersion originate primarily from the frequency-dependent polarization response of the substrate, rather than from the intrinsic frequency dependence of graphene alone.

The open circles in Fig.~\ref{fig:fig2} represent the numerical results obtained using COMSOL Multiphysics. Their colors indicate the propagation figure of merit, defined as
\begin{equation}
\mathrm{FOM}=
\frac{\operatorname{Re}(q)}
{|\operatorname{Im}(q)|}.
\label{eq:SPP_FOM}
\end{equation}

This dimensionless quantity characterizes the ratio between phase accumulation and attenuation during propagation. For a surface wave with complex propagation constant $q$, the intensity propagation length is defined as
\begin{equation}
L_{\mathrm{SPP}}=
\frac{1}{2|\operatorname{Im}(q)|},
\end{equation}
whereas the plasmon wavelength is
\begin{equation}
\lambda_{\mathrm{SPP}}=
\frac{2\pi}{\operatorname{Re}(q)}.
\end{equation}
Consequently, the propagation figure
of merit can also be written as
\begin{equation}
\mathrm{FOM}=
4\pi\frac{L_{\mathrm{SPP}}}
{\lambda_{\mathrm{SPP}}}.
\end{equation}

Larger FOM values therefore correspond to a greater number of plasmonic oscillation periods within the intensity attenuation length. As indicated by the color distribution of the numerical results in Fig.~\ref{fig:fig2}, the propagation quality varies considerably along the dispersion branches. In particular, a reduction in FOM is observed near the bending region of the lower branch, suggesting enhanced attenuation associated with the resonant dielectric response of SiO$_2$.

Overall, the agreement between the analytical dispersion curves and the full-wave numerical results supports applying the surface-conductivity description of graphene to the investigated structure. Comparison with the nondispersive reference system further shows that the optical phonon resonances of SiO$_2$ play a central role in shaping the dispersion and propagation characteristics of the hybrid plasmon--phonon surface modes.

\section{Localized Eigenmodes and Vortex States of a Graphene Ring}
\label{sec:antenna}

\subsection{Ring Geometry and Effective Eigenmode Formulation}
\label{sec:ring_eigenmodes}

We consider a planar graphene ring deposited on the surface of a dispersive SiO$_2$ substrate. The ring is characterized by its inner radius $a$, outer radius $b$, mean radius $R=(a+b)/2$, and width $w=b-a$. The graphene is described by the complex surface conductivity $\sigma(\omega)$, while the substrate is characterized by the frequency-dependent dielectric permittivity $\varepsilon_{\mathrm{SiO_2}}(\omega)$ introduced in Sec.~\ref{sec:materials}. The surrounding medium is air with $\varepsilon_1=1$. The same material parameters are used as in the analysis of propagating surface plasmon polaritons presented in Sec.~\ref{sec:spp_dispersion}.

The theoretical description follows the modal approach developed for circular ring-shaped resonators in Ref.~\onlinecite{He_OptExpress_2022}, with modifications required to account for the finite conductivity of graphene and the resonant dielectric response of the substrate. Unlike a conventional perfectly conducting microstrip resonator, the graphene ring supports strongly confined surface charge oscillations whose propagation characteristics are determined by the complex graphene conductivity and the surrounding dielectric environment.

Due to the rotational symmetry of the structure, the electromagnetic eigenfields can be classified according to the azimuthal index $m$ and radial index $n$. Using cylindrical coordinates $(\rho,\varphi,z)$, the angular dependence of the eigenfields can be represented as
\begin{equation}
\mathbf{E}_{mn}(\rho,\varphi,z)
=
\mathbf{e}_{mn}(\rho,z)e^{im\varphi},
\qquad m=0,\pm1,\pm2,\ldots.
\label{eq:ring_angular_field}
\end{equation}

For an ideal rotationally symmetric structure, the eigenmodes with azimuthal indices $+m$ and $-m$ are degenerate. Their linear combinations form standing-wave patterns proportional to $\cos(m\varphi)$ and $\sin(m\varphi)$. The corresponding angular distributions exhibit $2|m|$ alternating field lobes. In particular, the modes with $|m|=1$, $|m|=2$, and $|m|=3$ correspond to dipolar, quadrupolar, and hexapolar symmetries, respectively. The radial index $n$ distinguishes eigenstates with different radial distributions across the graphene annulus.

The resonant frequencies of the graphene ring are determined by the combined effects of azimuthal quantization, radial confinement, graphene conductivity, and the dispersive response of SiO$_2$. In contrast to a conventional dielectric ring resonator, the supported plasmonic modes may exhibit strong attenuation and pronounced frequency shifts near the optical phonon resonances of the substrate.

To establish a connection between the eigenmodes of the finite graphene ring and the propagating surface modes of the planar graphene--SiO$_2$ interface, we introduce an effective surface-wave approximation. Within this approximation, the modal amplitude $\Psi(\rho,\varphi)$ is assumed to satisfy a Helmholtz-type equation with the propagation constant $q_{\mathrm{SPP}}(\omega)$,
\begin{equation}
\left[
\nabla_{\parallel}^{2}
+
q_{\mathrm{SPP}}^{2}(\omega)
\right]\Psi(\rho,\varphi)=0,
\label{eq:ring_helmholtz}
\end{equation}
where $\nabla_{\parallel}^{2}$ is the two-dimensional Laplacian in the graphene plane. This approximation neglects the nonlocal electrostatic interaction associated with the finite lateral extent of the conducting sheet and is therefore intended to provide an effective estimate of the modal spectrum rather than an exact eigenvalue formulation.

Using separation of variables,
\begin{equation}
\Psi_{mn}(\rho,\varphi)
=
F_{mn}(\rho)e^{im\varphi},
\label{eq:ring_separation}
\end{equation}
the radial function satisfies
\begin{equation}
\frac{d^2F_{mn}}{d\rho^2}
+
\frac{1}{\rho}\frac{dF_{mn}}{d\rho}
+
\left[
q_{\mathrm{SPP}}^2(\omega)
-
\frac{m^2}{\rho^2}
\right]F_{mn}=0.
\label{eq:ring_radial_equation}
\end{equation}

The general solution is expressed in terms of Bessel functions of the first and second kinds,
\begin{equation}
F_{mn}(\rho)
=
A_{mn}J_m(q_{\mathrm{SPP}}\rho)
+
B_{mn}Y_m(q_{\mathrm{SPP}}\rho).
\label{eq:ring_bessel_solution}
\end{equation}

The finite radial extent of the graphene ring imposes boundary conditions at its inner and outer edges. Within a local hard-wall approximation, the normal component of the surface current vanishes at both edges,
\begin{equation}
J_{\rho}(a)=J_{\rho}(b)=0.
\label{eq:ring_current_boundary}
\end{equation}

Using the local constitutive relation $\mathbf{J}_s=\sigma(\omega)\mathbf{E}_{\parallel}$ and the electrostatic relation $\mathbf{E}_{\parallel}=-\nabla_{\parallel}\Psi$, these conditions reduce to
\begin{equation}
\left.
\frac{dF_{mn}}{d\rho}
\right|_{\rho=a}
=
\left.
\frac{dF_{mn}}{d\rho}
\right|_{\rho=b}
=0.
\label{eq:ring_neumann_boundary}
\end{equation}

Substitution of Eq.~\eqref{eq:ring_bessel_solution} into Eq.~\eqref{eq:ring_neumann_boundary} yields the characteristic equation
\begin{equation}
J_m'(q_{\mathrm{SPP}}a)Y_m'(q_{\mathrm{SPP}}b)
-
J_m'(q_{\mathrm{SPP}}b)Y_m'(q_{\mathrm{SPP}}a)
=0,
\label{eq:ring_characteristic}
\end{equation}
where the primes denote differentiation with respect to the arguments of the Bessel functions.

Equation~\eqref{eq:ring_characteristic} has the same mathematical structure as the characteristic equation for conventional ring-shaped resonators considered in Ref.~\onlinecite{He_OptExpress_2022}. The essential difference is that the effective propagation constant is now determined by the dispersive graphene--SiO$_2$ interface rather than by the wavenumber of a conventional dielectric cavity.

For a sufficiently narrow ring, $w/R\ll1$, and for the lowest radial family, the characteristic equation can be approximated by the azimuthal quantization condition
\begin{equation}
q_{\mathrm{SPP}}(\omega_{mn})R
\approx |m|.
\label{eq:ring_quantization}
\end{equation}

For the graphene ring considered here, the inner and outer radii are $a=35$~nm and $b=50$~nm, respectively, giving a mean radius $R=42.5$~nm and a width $w=15$~nm. The corresponding ratio $w/R\approx0.353$ indicates that the ring cannot be regarded as asymptotically narrow. Therefore, the azimuthal quantization condition in Eq.~\eqref{eq:ring_quantization} is used only as a qualitative guide to the ordering of the localized resonances. Quantitative predictions require accounting for the finite radial extent of the graphene annulus and the electrostatic coupling between its inner and outer boundaries, as discussed below.

\subsection{Quasistatic Integral Eigenvalue Problem}
\label{subsec:ring_integral}

A more general description of the graphene ring can be obtained in the quasistatic approximation,\cite{Wang_2012} which is appropriate when the characteristic dimensions of the resonator are much smaller than the free-space wavelength. This formulation retains the electrostatic interaction between different parts of the graphene annulus and does not require the introduction of an effective local surface-wave equation.

Let $\Phi(\rho,\varphi)$ denote the scalar electric potential on the graphene sheet. The tangential electric field and the induced surface current density are related by
\begin{equation}
\mathbf{E}_{\parallel}
=
-\nabla_{\parallel}\Phi,
\qquad
\mathbf{J}_s
=
-\sigma(\omega)\nabla_{\parallel}\Phi.
\label{eq:ring_surface_current}
\end{equation}

The continuity equation for the induced surface charge density $\rho_s$ reads
\begin{equation}
i\omega\rho_s
+
\nabla_{\parallel}\cdot\mathbf{J}_s
=0.
\label{eq:ring_continuity}
\end{equation}

Assuming a spatially uniform graphene conductivity within the annulus, Eqs.~\eqref{eq:ring_surface_current} and \eqref{eq:ring_continuity} give
\begin{equation}
\rho_s(\rho,\varphi)
=
\frac{\sigma(\omega)}{i\omega}
\nabla_{\parallel}^{2}\Phi(\rho,\varphi).
\label{eq:ring_charge_potential}
\end{equation}

For a graphene sheet located at the interface between two semi-infinite dielectric media, the quasistatic potential generated by the induced charge distribution is
\begin{equation}
\Phi(\mathbf{r})
=
\frac{1}
{2\pi\varepsilon_0[\varepsilon_1+\varepsilon_2(\omega)]}
\int_{\mathcal{A}}
\frac{\rho_s(\mathbf{r}')}{|\mathbf{r}-\mathbf{r}'|}
\,d^2\mathbf{r}',
\label{eq:ring_coulomb_potential}
\end{equation}
where $\mathcal{A}$ denotes the annular region occupied by graphene, $\varepsilon_1=1$, and $\varepsilon_2(\omega)=\varepsilon_{\mathrm{SiO_2}}(\omega)$. The dielectric response is assumed to be local and isotropic.

To account explicitly for the finite extent of the graphene annulus, we introduce the spatially dependent surface conductivity $\sigma(\mathbf{r},\omega)=\sigma(\omega)\chi_{\mathcal A}(\mathbf{r})$, where $\chi_{\mathcal A}$ is the characteristic function of the conducting region. Combining the continuity equation with the electrostatic relation between the surface current and potential yields the self-consistent integral equation
\begin{equation}
\Phi(\mathbf{r})=
\frac{1}{2\pi i\omega\varepsilon_0[\varepsilon_1+\varepsilon_2(\omega)]}
\int_{\mathbb R^2}
\frac{\nabla'_{\parallel}\cdot[\sigma(\mathbf{r}',\omega)\nabla'_{\parallel}\Phi(\mathbf{r}')]}{|\mathbf{r}-\mathbf{r}'|}
\,d^2\mathbf{r}'.
\label{eq:ring_integral_equation}
\end{equation}

The divergence form explicitly incorporates the spatial confinement of the conducting sheet. For a uniform conductivity within the annulus and vanishing normal current at its boundaries, the integral equation can be reduced to the corresponding form involving the in-plane Laplacian of the potential. The rotational symmetry then permits decomposition into independent azimuthal harmonics, leading to a radial integral eigenvalue problem for each angular index $m$.

Each azimuthal harmonic satisfies an independent radial integral equation,
\begin{equation}
\Phi_m(\rho)
=
\frac{\sigma(\omega)}
{i\omega\varepsilon_0[\varepsilon_1+\varepsilon_2(\omega)]}
\int_a^b
K_m(\rho,\rho')
\mathcal{L}_m\Phi_m(\rho')
\,\rho' d\rho',
\label{eq:ring_radial_integral}
\end{equation}
where
\begin{equation}
\mathcal{L}_m
=
\frac{d^2}{d\rho^2}
+
\frac{1}{\rho}\frac{d}{d\rho}
-
\frac{m^2}{\rho^2}
\label{eq:ring_radial_operator}
\end{equation}
and
\begin{equation}
K_m(\rho,\rho')
=
\frac{1}{2\pi}
\int_0^{2\pi}
\frac{e^{im\theta}}
{\sqrt{\rho^2+\rho'^2-2\rho\rho'\cos\theta}}
\,d\theta.
\label{eq:ring_coulomb_kernel}
\end{equation}

The integral equation~\eqref{eq:ring_radial_integral}, supplemented by the vanishing-normal-current conditions at the ring edges, defines the quasistatic eigenvalue problem for the graphene annulus. It accounts for the finite width of the ring and the nonlocal electrostatic coupling between different points of the conducting sheet, while retaining a local constitutive model for the graphene conductivity.

The geometric eigenvalues of the integral operator determine the allowed spatial charge distributions, whereas the frequency dependence of $\sigma(\omega)$ and $\varepsilon_{\mathrm{SiO_2}}(\omega)$ determines the corresponding complex resonance frequencies. In particular, the resonant response of the substrate can produce multiple spectral solutions associated with a given azimuthal symmetry, thereby providing a mechanism for the formation of distinct families of hybrid plasmon--phonon eigenmodes.

The integral formulation is employed here to establish the theoretical relationship between the geometry-dependent eigenfunctions and the frequency-dependent material response, whereas the complex eigenfrequencies and electromagnetic field distributions discussed below are obtained independently from full-wave finite-element calculations.

\subsection{Hybrid Plasmon--Phonon Eigenmodes}
\label{subsec:ring_hybrid}

The dispersion analysis of the planar graphene--SiO$_2$ interface revealed two separated spectral branches associated with the resonant dielectric response of the substrate. When the graphene sheet is patterned into a finite ring, the continuous surface-wave spectrum becomes discretized by the lateral geometry. Consequently, localized eigenmodes with the same azimuthal symmetry may occur in different frequency intervals corresponding to the lower and upper branches of the planar dispersion.

To distinguish these spectral families, we introduce an additional index $\nu=L,U$, where $L$ and $U$ denote the lower- and upper-frequency branches, respectively. The eigenmodes are then labeled by the azimuthal index $m$, radial index $n$, and branch index $\nu$, with complex eigenfrequencies
\begin{equation}
\widetilde{\omega}_{mn}^{(\nu)}
=
\omega_{mn}^{(\nu)}
+
i\gamma_{mn}^{(\nu)},
\qquad
\nu=L,U,
\label{eq:ring_complex_eigenfrequency}
\end{equation}
where $\gamma_{mn}^{(\nu)}>0$ represents the temporal decay rate of a passive eigenmode.

The corresponding modal quality factor is
\begin{equation}
Q_{mn}^{(\nu)}
=
\frac{\operatorname{Re}
\widetilde{\omega}_{mn}^{(\nu)}}
{2\operatorname{Im}
\widetilde{\omega}_{mn}^{(\nu)}}.
\label{eq:ring_quality_factor}
\end{equation}

The hybrid character of the eigenmodes originates from the simultaneous excitation of charge-carrier oscillations in graphene and the resonant dielectric polarization of SiO$_2$. The relative contributions of these mechanisms depend on the modal frequency, spatial distribution, and proximity to the optical phonon resonances of the substrate.

The azimuthal quantization condition provides a qualitative interpretation of the modal distribution between the two spectral branches. For a given ring radius, increasing the azimuthal order increases the characteristic in-plane propagation constant. Therefore, the resonance conditions for dipolar, quadrupolar, and hexapolar modes sample different regions of the planar SPP dispersion. However, the number of supported eigenmodes cannot be determined solely from the azimuthal quantization condition because the finite ring width, edge-induced charge accumulation, and frequency-dependent losses can substantially modify the eigenvalue spectrum.

The analysis of the complex eigenfrequencies and corresponding spatial field patterns allows the resonant states to be classified according to their azimuthal symmetry and spectral origin. Comparing modes with the same azimuthal index but different branch labels provides a direct means of investigating how the optical phonon response of SiO$_2$ modifies the localized plasmonic eigenstates of the graphene ring.

\begin{figure*}[!t]
\centering
\includegraphics[width=\linewidth]{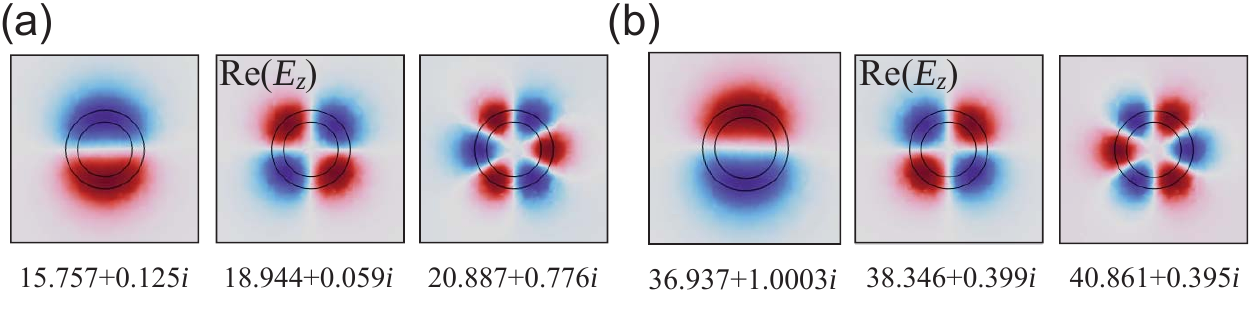}
\caption{Spatial distributions of the real part of the normal electric-field component, $\operatorname{Re}(E_z)$, for the eigenmodes of a graphene ring supported by a dispersive SiO$_2$ substrate. (a) Lower-frequency and (b) upper-frequency eigenmode families, each comprising dipolar ($m=1$), quadrupolar ($m=2$), and hexapolar ($m=3$) modes. The corresponding complex eigenfrequencies (in THz) are indicated below the field distributions. Red and blue regions represent opposite signs of $\operatorname{Re}(E_z)$. The inner and outer radii of the graphene ring are $a=35$~nm and $b=50$~nm, respectively, corresponding to a ring width of $15$~nm. The two eigenmode families are associated with the separated spectral branches of the graphene--SiO$_2$ surface plasmon--phonon polariton dispersion.}
\label{fig:fig3}
\end{figure*}

Figure~\ref{fig:fig3} presents the spatial distributions of the real part of the normal electric-field component, $\operatorname{Re}(E_z)$, for the eigenmodes of the graphene ring on the dispersive SiO$_2$ substrate. The numerical results reveal two distinct frequency groups associated with the lower and upper branches of the planar SPP dispersion shown in Fig.~\ref{fig:fig2}. Remarkably, both frequency groups support eigenmodes with the same sequence of azimuthal symmetries. The lower-frequency group comprises dipolar ($m=1$), quadrupolar ($m=2$), and hexapolar ($m=3$) modes with complex eigenfrequencies of $15.757+0.125i$, $18.944+0.059i$, and $20.887+0.776i$~THz, respectively. The corresponding eigenmodes of the upper-frequency group occur at $36.937+1.0003i$, $38.346+0.399i$, and $40.861+0.395i$~THz. The spatial field distributions exhibit two, four, and six alternating lobes, respectively, consistent with the characteristic azimuthal symmetries of dipolar, quadrupolar, and hexapolar oscillations.

The emergence of two spectrally separated families of localized eigenmodes is a distinctive feature of the graphene ring supported by a phonon-active dielectric substrate. Unlike a graphene resonator on a nondispersive dielectric, where the localized plasmonic response is governed primarily by the intrinsic graphene conductivity and geometric confinement, the present structure exhibits additional spectral complexity arising from the resonant dielectric response of SiO$_2$. The appearance of dipolar, quadrupolar, and hexapolar modes in both frequency groups indicates that the same azimuthal field symmetries can be realized at substantially different frequencies. This behavior is consistent with the hybrid plasmon--phonon character of the underlying surface waves and demonstrates that substrate dispersion provides an additional mechanism for controlling the spectral positions and attenuation of localized graphene resonances.

The complex eigenfrequencies also reveal pronounced differences in modal damping. Using $Q=\operatorname{Re}(\widetilde{f})/[2\operatorname{Im}(\widetilde{f})]$ for the adopted $\exp(+i\omega t)$ convention, the estimated quality factors of the lower-frequency dipolar, quadrupolar, and hexapolar modes are approximately $63.0$, $160.5$, and $13.5$, respectively. The corresponding upper-frequency modes exhibit quality factors of approximately $18.5$, $48.1$, and $51.7$. Notably, the lower-frequency quadrupolar mode possesses the highest quality factor among the identified eigenstates, whereas the upper-frequency quadrupolar and hexapolar modes exhibit comparable damping rates. These results indicate that modal attenuation depends not only on the azimuthal order but also on the spectral position relative to the resonant dielectric response of SiO$_2$. Quantitatively decomposing the losses into graphene absorption, substrate dissipation, and radiation leakage would require additional analysis of the corresponding electromagnetic energy balance.

It should be noted that the calculated complex eigenfrequencies are sensitive to the numerical eigenvalue solver, its convergence settings, and the treatment of frequency-dependent material parameters, particularly the dispersive response of SiO$_2$. Consequently, the reported eigenfrequencies and the corresponding quality factors should be regarded as approximate estimates rather than quantitatively definitive predictions. The present eigenmode analysis is therefore intended primarily to provide a qualitative description of the resonant spectrum, including the emergence of two spectrally separated families of multipolar modes and their characteristic spatial symmetries. A quantitatively accurate determination of the complex resonance frequencies and modal damping would require a dedicated treatment of the nonlinear dispersive eigenvalue problem and further convergence analysis.

\subsection{Construction of Quadrupolar Vortex States}
\label{subsec:ring_vortex}

The quadrupolar eigenmodes identified in both spectral branches of the graphene--SiO$_2$ structure provide a basis for constructing localized electromagnetic vortex states with well-defined azimuthal phase winding. Following the modal superposition principle developed for circular ring-shaped resonators in Ref.~\onlinecite{He_OptExpress_2022}, an azimuthally traveling-wave state can be formed by combining two orthogonal standing-wave eigenmodes with equal amplitudes and a relative phase shift of $\pm\pi/2$. In a rotationally symmetric structure, these orthogonal components belong to the same degenerate azimuthal mode family and differ only in their angular orientation.

For the quadrupolar eigenmodes ($m=2$), the two orthogonal standing-wave solutions can be represented by angular dependences $\cos(2\varphi)$ and $\sin(2\varphi)$. Their coherent superposition yields
\begin{equation}
\Psi_{\pm2}^{(\nu)}(\rho,\varphi)=F_2^{(\nu)}(\rho)\left[\cos(2\varphi)\pm i\sin(2\varphi)\right]=F_2^{(\nu)}(\rho)e^{\pm i2\varphi},
\label{eq:ring_vortex}
\end{equation}
where $F_2^{(\nu)}(\rho)$ describes the radial field distribution and $\nu=L,U$ distinguishes the lower- and upper-frequency eigenmode families. The resulting states exhibit opposite azimuthal phase winding, characterized by scalar topological charges $l=\pm2$. Accordingly, the complex field phase accumulates $\pm4\pi$ upon one complete revolution around the symmetry axis.

\begin{figure*}[!t]
\centering
\includegraphics[width=\textwidth]{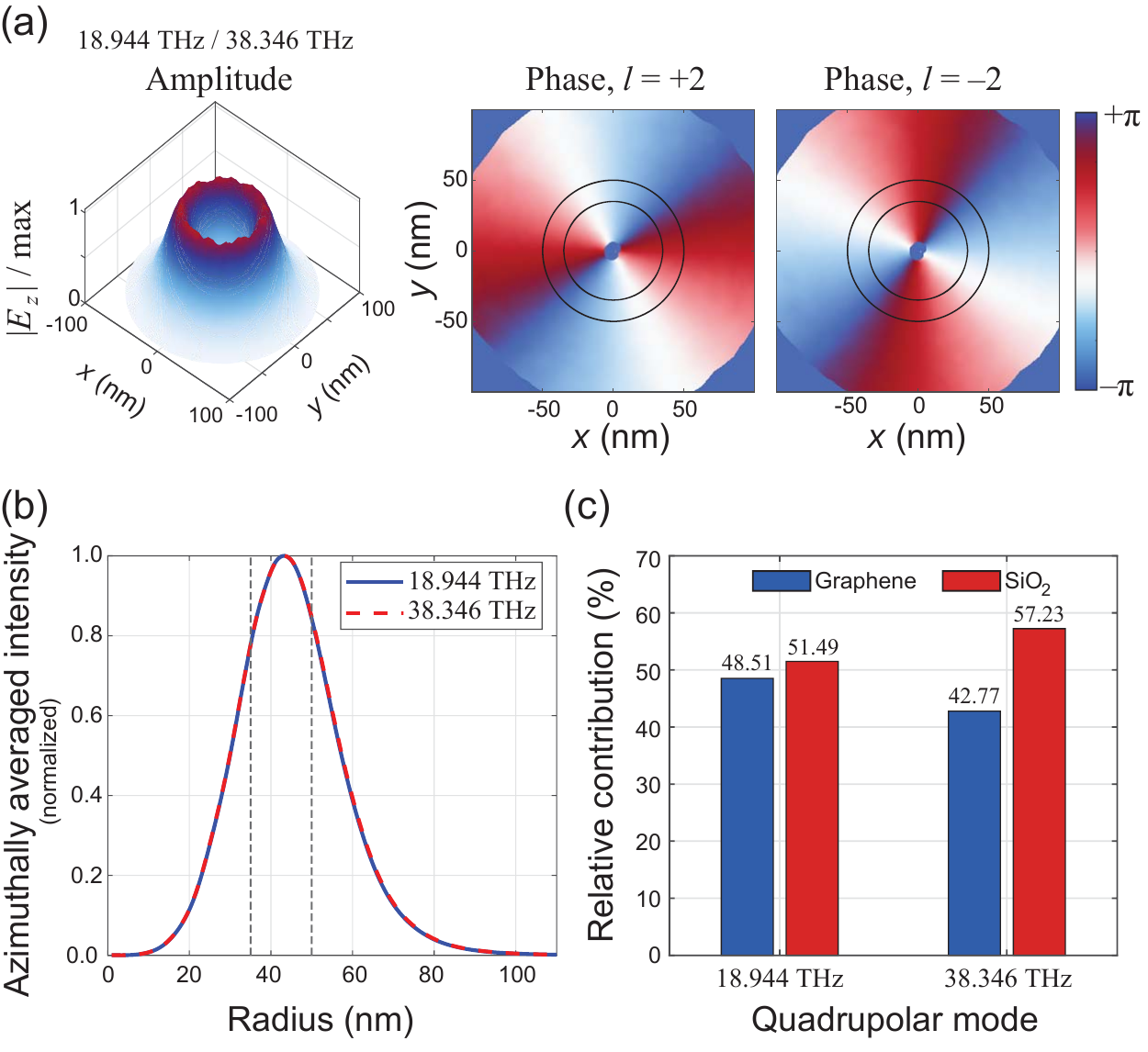}
\caption{Quadrupolar vortex states and material-loss characteristics of a graphene ring supported by a dispersive SiO$_2$ substrate. (a) Representative three-dimensional distribution of the normalized electric-field amplitude, $|E_z|/\max|E_z|$, and phase distributions of the vortex states with topological charges $l=+2$ and $l=-2$, constructed through the coherent superposition of orthogonal quadrupolar field patterns. The nearly identical normalized spatial profiles of the quadrupolar resonances at $18.944$ and $38.346$~THz allow the same representative vortex-field distributions to be shown for both frequency branches. (b) Comparison of the azimuthally averaged radial intensity profiles of the two quadrupolar modes, normalized independently to their respective maxima. The vertical dashed lines indicate the inner ($a=35$~nm) and outer ($b=50$~nm) radii of the graphene ring. (c) Relative contributions of graphene and SiO$_2$ to the integrated electromagnetic loss-density quantity, obtained from the corresponding eigenmode solutions.}
\label{fig:fig4}
\end{figure*}

To construct these vortex states numerically, we use the complex distributions of the normal electric-field component, $E_z(x,y)$, obtained from the full-wave eigenfrequency calculations in COMSOL Multiphysics. For each quadrupolar resonance, the calculated field is combined with a spatially rotated copy of itself. A rotation by $\alpha=\pi/(2m)=\pi/4$ transforms the angular dependence $\cos(2\varphi)$ into an orthogonal quadrupolar distribution proportional to $\sin(2\varphi)$. The two components are then superimposed with a relative phase shift of $\pm\pi/2$, giving
\begin{equation}
E_{z,\pm}^{(\nu)}(x,y)=E_z^{(\nu)}(x,y)\pm iE_{z,\mathrm{rot}}^{(\nu)}(x,y),
\label{eq:ring_vortex_superposition}
\end{equation}
where $E_{z,\mathrm{rot}}^{(\nu)}$ denotes the original complex eigenfield distribution rotated by $45^\circ$ about the center of the graphene ring. This numerical procedure constructs traveling azimuthal field patterns from the calculated standing-wave eigenmodes without requiring an additional electromagnetic excitation simulation.

The procedure is applied independently to the lower-frequency quadrupolar eigenmode at $\widetilde{f}_{2}^{(L)}=18.944+0.059i$~THz and the upper-frequency quadrupolar eigenmode at $\widetilde{f}_{2}^{(U)}=38.346+0.399i$~THz. Figure~\ref{fig:fig4}(a) illustrates the resulting vortex-field structure. The normalized electric-field amplitude exhibits an approximately rotationally symmetric annular distribution with a pronounced central minimum, whereas the phase maps show opposite azimuthal phase winding for $l=+2$ and $l=-2$. The topological charges are verified by evaluating the phase accumulation along a closed circular contour of radius $R=42.5$~nm,
\begin{equation}
l=\frac{1}{2\pi}\oint_C\nabla_{\parallel}\arg(E_z)\cdot d\mathbf{l},
\label{eq:ring_topological_charge}
\end{equation}
which gives $l=\pm2$ for the corresponding vortex states. The nearly identical normalized vortex-field distributions obtained from the two quadrupolar resonances allow representative amplitude and phase maps to be shown for both frequency branches.

A remarkable feature of the two quadrupolar resonances is the near identity of their spatial field distributions despite their considerable spectral separation. A direct comparison of the complex eigenfields obtained at $18.944$ and $38.346$~THz yields a normalized modal overlap of approximately $0.9998$, with a phase-aligned relative $L^2$ difference of about $2.1\%$. Figure~\ref{fig:fig4}(b) compares the azimuthally averaged radial intensity profiles of the two eigenmodes. The curves nearly coincide, indicating that the spatial localization of the quadrupolar fields is largely preserved between the lower- and upper-frequency spectral branches. This behavior is consistent with the quasistatic description of the graphene ring, in which the spatial eigenfunctions are primarily governed by the resonator geometry, whereas the frequency-dependent electromagnetic response of graphene and SiO$_2$ determines the corresponding resonance frequencies.

Despite their nearly identical normalized spatial distributions, the two quadrupolar eigenmodes exhibit different estimated attenuation characteristics. The calculated quality factors suggest a more weakly damped resonance in the lower-frequency spectral branch than in the upper-frequency branch. The estimated quality factors are approximately $Q_L=160.5$ and $Q_U=48.1$ for the lower- and upper-frequency resonances, respectively (these estimates are subject to the numerical limitations of the dispersive eigenfrequency analysis discussed above). To examine the corresponding material contributions, the total power dissipation density was integrated separately over the graphene layer and the SiO$_2$ substrate. The relative contribution of each material was calculated by normalizing its integrated value to the sum of the two material contributions. As shown in Fig.~\ref{fig:fig4}(c), the lower-frequency quadrupolar mode exhibits relative contributions of $48.51\%$ from graphene and $51.49\%$ from SiO$_2$, whereas the corresponding values for the upper-frequency mode are $42.77\%$ and $57.23\%$, respectively. Thus, the upper-frequency resonance exhibits a larger relative contribution from the substrate despite retaining almost the same spatial field structure. These quantities are evaluated from the eigenmode solutions and are used here as comparative indicators of the material-loss distribution rather than as a complete decomposition of the resonant decay rates.

The results demonstrate that two spectrally separated quadrupolar eigenmodes of the same graphene ring provide a basis for constructing localized vortex states with identical absolute topological charges but different modal attenuation characteristics. The normalized spatial profiles of these states remain nearly unchanged, whereas their complex resonance frequencies and relative material-loss indicators differ substantially. This behavior is consistent with a picture in which the ring geometry primarily determines the spatial symmetry of the confined plasmonic oscillations, while the dispersive response of graphene and SiO$_2$ governs their spectral properties. The constructed vortex states establish the feasibility of a multiband near-field vortex configuration; its direct excitation by external sources remains to be demonstrated.

\section{Conclusions}
\label{sec:concl}

We have investigated the formation of hybrid plasmon--phonon eigenmodes and localized vortex states in a graphene ring supported by a dispersive SiO$_2$ substrate. An analytical description of TM-polarized surface plasmon polaritons at the planar graphene--SiO$_2$ interface was developed and validated against full-wave numerical simulations. The calculations reveal two separated spectral branches associated with the resonant dielectric response of the substrate. Comparison with a nondispersive dielectric reference confirms the important role of the infrared-active optical phonons of SiO$_2$ in determining the surface-wave dispersion and propagation characteristics.

The analysis was extended to a finite graphene ring with inner and outer radii of $35$ and $50$~nm, respectively. Full-wave eigenfrequency calculations identify dipolar, quadrupolar, and hexapolar resonances in both spectral regions. In particular, the quadrupolar modes at $18.944$ and $38.346$~THz exhibit nearly identical normalized spatial distributions of the normal electric-field component, with a calculated complex field overlap of approximately $0.9998$. Nevertheless, the corresponding estimated quality factors differ substantially, reaching approximately $160.5$ and $48.1$, respectively. Analysis of the integrated electromagnetic loss-density quantity further reveals a change in the relative contribution of SiO$_2$ from $51.49\%$ to $57.23\%$ between the lower- and upper-frequency quadrupolar resonances. These results indicate that closely similar spatial eigenfield patterns can occur at widely separated frequencies while exhibiting different attenuation characteristics.

Using the calculated complex eigenfields, localized quadrupolar vortex states were constructed through coherent superposition with spatially rotated field distributions. The resulting states exhibit opposite azimuthal phase winding with scalar topological charges $l=\pm2$. The persistence of the quadrupolar spatial structure across the two spectral branches provides a basis for generating similar structured near fields at substantially different infrared frequencies without changing the resonator geometry. The demonstrated combination of geometric confinement and substrate-induced plasmon--phonon hybridization offers a route toward compact multiband platforms for controlling the phase structure of localized electromagnetic fields.

%-------

\section*{Acknowledgments}
V. R. Tuz is grateful for the support from Jilin University, China.

\section*{AUTHOR DECLARATIONS}

\subsection*{Conflict of Interest}
The authors have no conflicts to disclose.

\subsection*{Author Contributions}
\textbf{Y. Guo:} Data curation (equal); Formal analysis (equal); Investigation (lead); Writing – original draft (supporting). \textbf{O. Demianyk:} Data curation (equal); Investigation (supporting); Writing – original draft (supporting). \textbf{V. R. Tuz:} Conceptualization (lead); Methodology (lead); Formal analysis (equal); Validation (lead); Writing – original draft (lead); Supervision (lead); Funding acquisition (lead).

\bigskip

\section*{Data Availability}
The data that support the findings of this study are available from the corresponding author upon a reasonable request.

\bibliography{graphene_patch}

\end{document}